\documentclass[a4paper,twocolumn,showpacs,prl]{revtex4}
\usepackage{epsfig}
\usepackage{epstopdf}
\usepackage{bm} 
\usepackage{amsmath, amssymb} 
\usepackage{dsfont} 

\begin{document}

\title{Universal non-linear conductivity near to an itinerant-electron
  ferromagnetic quantum critical point}

\author{P. M. Hogan}
\affiliation{School of Physics and Astronomy, University of St
  Andrews, North Haugh, St Andrews, KY16~9SS, UK}

\author{A. G. Green}
\affiliation{School of Physics and Astronomy, University of St
  Andrews, North Haugh, St Andrews, KY16~9SS, UK}

\date{\today}

\begin{abstract}
We study the conductivity in itinerant-electron systems near to
a magnetic quantum critical point. We show that, for a class of
geometries, the universal power-law
dependence of resistivity upon temperature may be reflected in a universal
non-linear conductivity; when a strong electric field is applied, the
resulting current has a universal power-law dependence upon the
applied electric field. For a system with thermal equilibrium
current proportional to $T^\alpha$ and dynamical exponent $z$, we
find a non-linear resistivity proportional to $E^{\frac{z-1}{z(1+\alpha)-1}}$.
\end{abstract}

\pacs{72.10.Di, 72.90.+y, 75.30.Kz, 75.40.-s}

\maketitle

The notion of quantum criticality provides one of the few unifying
theoretical principles of strongly correlated
electrons~\cite{Sachdev, Sondhi97, Coleman05}. It describes
a range of phenomena in systems that are near to continuous,
zero-temperature phase transitions; phase transitions that are driven
by quantum rather than thermal fluctuations. In thermal
equilibrium, quantum critical systems show characteristic spatial and
temporal scaling in their response to external probes. For example, the
conductivity of an itinerant-electron system near to a magnetic
quantum phase transition has a power-law dependence upon
temperature~\cite{Moriya, Hertz76, Millis}.

The behaviour of quantum critical systems out of thermal equilibrium
has begun to attract growing attention over the past few years.
Near to a quantum phase
transition all of the intrinsic energy scales of a system, other than the Fermi energy,
renormalize to zero. In thermal equilibrium, the only remaining energy
scale is the temperature itself. Because of this, quantum critical
systems are particularly susceptible to being driven out of
equilibrium by external probes. In certain
situations the universal temporal scaling near to the quantum critical
point may reveal itself in universal features of the steady-state
adopted out of equilibrium; the out-of-equilibrium state being largely
determined by a system's dynamics.

Several recent works have addressed the question of whether
universality persists when a quantum critical system is driven out of
thermal equilibrium by the application of a strong electric field. In
particular, Refs.~\cite{Dalidovich04} and~\cite{Green05, Green06}
considered two-dimensional superconductor-insulator
transitions~\cite{Cha91, Damle97}, the former in the case where the
quantum dynamics and phase transition were controlled by coupling to a
Caldeira-Leggett bath and the latter in the case of intrinsic
superconducting dynamics. These systems can indeed display
universality out of equilibrium~\cite{Fenton05} in both their current
response~\cite{Dalidovich04,Green05} and their current noise
statistics\cite{Green06}. The triumph of
Refs.~\cite{Dalidovich04} and~\cite{Green05} was to provide
field-theoretical derivations of the scaling predicted by na\"{\i}ve
dimensional analysis.  Numerical studies of related one-dimensional
systems produced similar results~\cite{Aoki}.

Whilst these works provide interesting proofs of principle --- and
indeed, may yet be compared with experiment --- most quantum
critical systems that are studied experimentally are of a rather
different type. The critical modes at the superconductor-insulator
transition are charged and couple directly to the electric field. A
more typical situation has critical modes without a charge --- often
magnetic --- which affect transport by scattering from
electrons. Here we address the question of whether universal
non-linear response in transport occurs in this more general
setting.

We find that given certain conditions on size and geometry, quantum critical
itinerant magnets show a universal
non-linear current response. For a long, narrow sample, with an electric
field applied along its length, we predict a universal non-linear
scaling of current with field given by
\begin{equation}
j \propto E^{\frac{z-1}{z(1+\alpha)-1}},
\label{firstEq}
\end{equation}
where the thermal equilibrium resistivity is proportional to $T^\alpha$ and $z$ is the dynamical exponent. In the case of the Moriya-Hertz-Millis~\cite{Moriya, Hertz76, Millis} model of the critical ferromagnet,
$\alpha=(d+z-1)/z$.
Provided that certain constraints upon the dimension of the system are satisfied, this result does not depend further upon the system dimensions.  In the
following, we will take some time to discuss this matter and
compare our results to those of related works.

We hope that these results will provide an alternative experimental
window upon quantum criticality. Despite its successes, the theory of
itinerant-electron quantum criticality has some puzzling problems; although
power law dependencies upon temperature are seen experimentally, there is often a
discrepancy between the observed and predicted powers in transport.
Non-linear response may help to resolve this issue by providing two consistency checks: whether the equilibrium exponents are consistent with the out-of-equilibrium exponents through Eq.(\ref{firstEq}); and whether the out-of-equilibrium exponent is consistent with the Moriya-Hertz-Millis theory.

Our paper is outlined as follows: we begin in Section 1 with a
general description of our scheme, paying particular attention to
matters of geometry and heat flows within the system. This will enable
a heuristic derivation of our main results and a detailed comparison
with the complementary work of Mitra {\it et al.}~\cite{Mitra06}. In Section 2,  we will
begin with a survey of the Boltzmann treatment of the linear response of the itinerant-electron
quantum critical system in thermal equilibrium. This will
allow us to introduce some notation and familiarize the reader with
its application in this context. We follow this in Section 3 by
applying the Boltzmann transport formalism to the out of equilibrium
system. This section contains a formal derivation of our main
results. Finally, in Section 4, we turn to a discussion of the
limitations of our analysis and of the prospects for seeing the
effects that we predict in experiment.

\section{General Scheme}

\noindent
{\it Geometry}: We consider a long, narrow, quantum-critical, itinerant
electron system
with an electric field applied along its length. The system is longer
than its transport length so that an electron traversing the sample
scatters from magnons many times. The system must also be wide enough that
it displays bulk behaviour, but narrow enough that heat generated
within the sample can be transported to the boundary.

\noindent
{\it Critical Fields}: Starting from some low base temperature, $T_0$,
and gradually increasing
the electric field one may anticipate two fields at which the response
may become non-linear:

\noindent
i. When the energy gained by an electron from the electric field
between scattering events exceeds the temperature;
$$
E_1 \sim \frac{T_0}{l_{\textrm{tr}}},
$$
where $l_{\textrm{tr}}$ is the transport scattering length.

\noindent
ii. When the Joule heating rate exceeds the rate at which heat may be
transported from the sample by a transverse heat flow;
\begin{eqnarray*}
E_2^2 \sigma \sim \kappa T_0/ W^2
\;\;\Rightarrow\;\;
E_2 \sim \frac{T_0}{W} \sqrt{l_{\textrm{th}}/l_{\textrm{tr}}}
\end{eqnarray*}
where $\sigma$ and $\kappa$ are the electrical and thermal
conductivities. The latter result has been obtained using $\kappa/ \sigma T_0 =
l_{\textrm{th}}/l_{\textrm{tr}}$.  $l_{\textrm{th}}$ is the thermal scattering length and $W$ is the
sample width.

In this work, we will be primarily concerned with the former
case. In order for a system to be in this regime, we require that
$E_1 \ll E_2$, so that we hit the field $E_1$ first when increasing the electric field from zero; {\it i.e.} we
require that $W \ll\sqrt{l_{\textrm{tr}} l_{\textrm{th}}}$. In addition, for the
system to exhibit bulk behaviour requires that it be wider than its
correlation length, $W \gg l_{\textrm{th}}$. Combining these two conditions upon
the sample width yields
\begin{equation}
l_{\textrm{th}} \ll W \ll \sqrt{l_{\textrm{tr}} l_{\textrm{th}}}.
\label{Regime}
\end{equation}
Due to additional angular factors, the transport length is
substantially greater than the thermal scattering length, $l_{\textrm{tr}}\gg
l_{\textrm{th}}$, so that there is a large window of sample widths over which
the type of non-linear response that we envisage can occur. In the
high-temperature limit (with $T\ll \epsilon_{\textrm{F}}$ nevertheless) in which
experimental investigations of itinerant electron quantum criticality
are usually carried out,
$l_{\textrm{tr}} \sim l_{\textrm{th}}/\theta^2$
where $\theta \sim q/k_{\textrm{F}} \sim (T/\epsilon_{\textrm{F}})^{1/z}$ is the angle of scattering.

This regime is somewhat delicately balanced between
the macro- and microscopic. In a truly macroscopic sample where $W
\rightarrow \infty$, non-linearity always occurs due to the failure to
conduct away excess Joule heat. In our case, the transverse size of
the system must be small enough that $W  \ll \sqrt{l_{\textrm{tr}}
  l_{\textrm{th}}}$, but the system inherits its behaviour from macroscopic
equilibrium properties since it is larger than the correlation
length\cite{footnote1}.
If the above constraints are satisfied, the non-linear transport properties depend only upon bulk properties and are independent of the dimensions of the system.

\vspace{0.1in}
\noindent
{\it Thermal Coupling}: Determining the non-linear response requires keeping
careful track of the various heat flows. We consider a simplified
scheme of thermal couplings in our sample:

\noindent
i. The electrons couple to a heat sink at the boundaries of the sample
and scatter from magnons. We do not consider electron-electron
scattering since this is higher order in temperature or electric field
than electron-magnon scattering and so sub-leading at low temperatures
and fields.

\noindent
ii. The magnons may scatter both from one another and from the
electrons. We do not consider coupling between magnons and the heat
sink. Our reason is that magnon-phonon relaxation is higher order in
temperature or field than magnon-electron scattering and therefore
weaker at low temperatures and field.

\vspace{0.1in}
\noindent
{\it Heuristic Treatment}: Given these descriptions of the geometry of
our system and the various
microscopic couplings, we are now in a position to give a heuristic
derivation of our main results. Heat enters the system via Joule
heating and ultimately leaves through a transverse heat flow
maintained by a transverse variation in temperature. In the absence of
scattering between electrons, this energy must pass through the
magnons: Joule heating pumps energy into higher moments of the
electron distribution. This is ultimately carried away by a transverse heat flow maintained by a gradient in the symmetrical part of the electron distribution. Energy can only pass into the symmetrical part of
the distribution due to scattering {\it via} magnons. In leading
approximation, the magnons are raised to an effective temperature
$T_{\textrm{eff}}(E)$ determined by a balance between the Joule heating rate
and the rate at which the magnons at $T_{\textrm{eff}}(E)$ lose energy to the
electrons at $T_0 \sim 0$. Equating these two rates of
change of energy leads to a self-consistency equation for $T_{\textrm{eff}}$.

In the high-temperature limit, the scattering time $\tau$, the transport scattering time $\tau_{\textrm{tr}}$ and the magnon energy decay rate $d\xi/dt$ are related as follows:
\begin{eqnarray*}
\frac{1}{\tau_{\textrm{tr}}}
& \sim &
\frac{T^{2/z}}{\tau},
\\
\frac{d {\cal E}}{dt}
& \sim &
\frac{T^2}{\tau}.
\end{eqnarray*}
Using these relations for a system whose thermal equilibrium resistivity scales as $T^\alpha$, we find
\begin{eqnarray}
\hbox{Joule Heating}
& \propto&
\sigma E^2
\propto
E^2 T_{\textrm{eff}}^{-\alpha},
\nonumber\\
\hbox{Energy Relaxation}
& \propto&
T_{\textrm{eff}}^{2 (1-1/z+\alpha)},
\nonumber\\
\frac{d {\cal E}}{dt}
&=&
\sigma E^2.
\label{EnergyBalance}
\end{eqnarray}
By equating these two rates leads we deduce that
$T_{\textrm{eff}} \propto E^{z/(z(1+\alpha)-1)}$
and
$j \propto E^{(z-1)/(z(1+\alpha)-1)}$.
The following sections will flesh out these ideas with particular reference to the Moriya-Hertz-Millis~\cite{Moriya, Hertz76, Millis} model of the critical ferromagnet.

\vspace{0.1in}
\noindent
{\it Comparison with Mitra et al.}: A recent work of Mitra {\it et al} has
considered the same system as studied here, but in a different
geometry. The results obtained in Ref.\cite{Mitra06} are different from ours
because of this geometry. In order to allay any confusion,
it is worth spending a moment to note the main distinction between our
two works. Mitra {\it et al}\cite{Mitra06} consider an itinerant
electron system with essentially two-dimensional geometry and an
electric field applied in the third short direction. In this case, an
electron traversing the sample from one lead to another does not
scatter appreciably from magnons--- the electron distribution is
determined to leading order by the distributions in the leads and may
be written directly in terms of them using a Keldysh formalism. Mitra
{\it et al} present an appealing derivation of this zeroth order
distribution and show, using a renormalization group analysis, that an
effective temperature proportional to the applied voltage results. In
our case, by contrast, an electron traversing the sample between the
two leads scatters many times off magnons and the electron
distribution must be calculated self-consistently from the
start\cite{footnote2}.

In the rest of this paper, we will spend some time fleshing out the
mathematical details of this general scheme. We begin in the next
section by reviewing the Boltzmann approach to thermal equilibrium
transport in quantum critical metals.

\section{Boltzmann Approach in Thermal Equilibrium}
We will use Boltzmann transport techniques to analyse
the out-of-equilibrium response of a quantum critical system to an electric
field. Although this approach is familiar in other contexts,
itinerant-electron quantum-critical transport is usually analysed by other means.
Therefore, in this section, we will spend a little
time summarising quantum critical transport in thermal equilibrium and
how this may be described using a Boltzmann equation approach. This
exposition will also serve as a useful way of defining the notation
that we will use later in our analysis of the non-linear response.

Our first step will be to describe the thermal-equilibrium magnon
propagator. We follow this by writing down the electron Boltzmann
equation and construct its linear response solution.
Finally, we quote a number of relaxation rates that will useful in our
non-equilibrium analysis. The details of the calculation of these
within a Boltzmann framework is somewhat similar to that of the
relaxation rates due to phonon scattering. We sketch these calculations
in  Appendix A.

\subsection{Magnon Propagator}
We work within the Moriya-Hertz-Millis~\cite{Moriya, Hertz76, Millis} approach to itinerant electron
quantum criticality. The bosonic, magnetic critical modes ---
the magnons --- are treated separately from the electrons (although they
are, of course, made from electrons). The effects of scattering
between the magnons and electrons are treated self-consistently; the
magnon dynamics being determined by
Landau damping and the electronic transport being determined
by scattering from the magnon fluctuations.

The first step in the Moriya-Hertz-Millis approach
to itinerant electron quantum criticality, is to determine
the critical properties of the magnons. These critical properties are
the combined result of the magnons' self-interaction and their
overdamped dynamics due to Landau damping. The simplest way to do this
is through the
self-consistent renormalization group\cite{Moriya}. Alternatively, one
may use a more rigorous application of the renormalization
group\cite{Hertz76,Millis} in
order to obtain essentially the same results. In either case, the
critical magnon propagator takes on the following form in the
equilibrium quantum critical state:
\begin{equation}
D^{\textrm{R}}({\bm{q}}, \omega)
=
\left[
\textrm{i}\frac{|\omega|}{\Gamma_{\bm{q}}} +{\bm{q}}^2 + r(T)
\right]^{-1},
\label{MagnonPropagator}
\end{equation}
where $\Gamma_{\bm{q}}$ describes the Landau damping. $\Gamma_{\bm{q}}$
is proportional to $|{\bm{q}}|$ in
the ferromagnet and constant in an antiferromagnet
 (or $\Gamma_{\bm{q}} = \Gamma |{\bm{q}}|^{z-2}$ in
general). The magnon gap $r(T)$ takes on characteristic power-law
forms in temperature in the quantum critical system;
\begin{equation}
r(T) \propto
T^{\frac{d+z-2}{z}},
\label{rofT}
\end{equation}
 where $d$ is the dimension and $z$ is the
dynamical exponent ($z=3$ in the ferromagnet and $2$ in the
antiferromagnet).
Through most of our subsequent analysis, we shall concentrate upon the situation in three dimensions. This is readily extended to other dimensions.
The overdamping of magnons has important
consequences. Unlike their phonon counterparts, magnon excitations do
not have a well-defined energy for a particular wave-vector. This does
not have enormous consequences for a Boltzmann analysis in thermal
equilibrium, but it does necessitate modification of the magnon
Boltzmann equation when we consider the out-of-equilibrium situation.

\subsection{The Boltzmann Equation}
The electronic Boltzmann equation for scattering from an uncharged
auxiliary mode may be written
\begin{equation}
\begin{split}
&\left[ \partial_t +(e{\bm{E}}/\hbar)\cdot\partial_{{\bm{k}}} \right]f_{{\bm{k}}}
\\&=
-\int \frac{d{\bm{q}}}{(2 \pi)^3} \;
\left[
\gamma_{{\bm{k}}{\bm{q}}} f_{{\bm{k}}} \left( 1 - f_{{\bm{q}}} \right)
-
\gamma_{{\bm{q}}{\bm{k}}} f_{{\bm{q}}} \left( 1 - f_{{\bm{k}}} \right)
\right].
\label{Electron_Boltzmann1}
\end{split}
\end{equation}
The matrices $\gamma_{{\bm{k}}{\bm{q}}}$ describe scattering from the
auxiliary modes--- magnons in our case. Quite generally, for scattering
from auxiliary modes that have a thermal distribution, the
scattering matrices satisfy the detailed balance relationship
\begin{equation}
\gamma_{{\bm{k}}{\bm{q}}}
=\gamma_{{\bm{q}}{\bm{k}}}
\exp[ (\epsilon_{{\bm{k}}}-\epsilon_{{\bm{q}}} )/T].
\label{DetailedBalance}
\end{equation}
In the case of magnon scattering, the scattering matrices take the form
\begin{equation}
\gamma_{{\bm{p}}{\bm{q}}}
=
|g_{{\bm{q}}-{\bm{p}}}|^2
\left[
\begin{array}{l}
\left[
1+n(\epsilon_{\bm{p}}-\epsilon_{\bm{q}}) \right]
\rho({\bm{p}}-{\bm{q}}, \epsilon_{\bm{p}}-\epsilon_{\bm{q}})
\\
\;\;\;\;\;\;\;\;
+n(\epsilon_{\bm{q}}-\epsilon_{\bm{p}})
\rho({\bm{q}}-{\bm{p}}, \epsilon_{\bm{q}}-\epsilon_{\bm{p}})
\end{array}
\right]
\label{MagnonGamma}
\end{equation}
%
where $g_{{\bm{q}}-{\bm{p}}}$ is the matrix element for electron-magnon scattering and
$\rho({\bm{q}}, \omega)$ is the magnon spectral function. In antiferromagnets, the matrix element $g_{{\bm{q}}-{\bm{p}}}$ has significant momentum dependence, with scattering hot spots corresponding to resonance of the magnetic ordering wave-vector with the Fermi surface. For simplicity, we restrict our analysis to the case of ferromagnets or long wavelength helimagnets where the momentum dependence of $g_{{\bm{q}}-{\bm{p}}}$ is weak and can be neglected. The magnon spectral function is given by
\begin{equation}
\rho({\bm{q}}, \omega)
=
-\frac{1}{\pi}
{\cal I} m D^{\textrm{R}}({\bm{q}}, \omega)
=
\frac{\omega/\Gamma_{\bm{q}}}
{(r+{\bm{q}}^2)^2+ (\omega/\Gamma_{\bm{q}})^2}.
\label{MagnonSpectral}
\end{equation}
It is
determined by the magnon
propagator given in Eq.~(\ref{MagnonPropagator}). It contains all of
the information about how dynamics is incorporated into the critical
behaviour through the relative scaling of frequency and momentum:
$\omega \sim q^z \sim q^2 \Gamma_{\bm{q}}$. In what
follows, it will prove very useful to work with the general form of
the Boltzmann equation (\ref{Electron_Boltzmann1}) rather than the form obtained
after explicit substitution of $\gamma_{{\bm{p}}{\bm{q}}}$.

\subsection{Linear Response Solution of the Boltzmann Equation}
The generic notation of Eq.~(\ref{Electron_Boltzmann1}) allows us to
construct a formal linear response solution of the Boltzmann equation both in thermal equilibrium and,
ultimately, out of thermal equilibrium. In order to orient ourselves
for the latter more involved calculation, let us first construct the
conventional linear response solution with this general notation.
Identifying~\cite{Green06}
\begin{equation}
\mathsf{M}_{{\bm{k}}{\bm{q}}}
=
\frac{ \gamma_{{\bm{q}}{\bm{k}}}   }{\gamma_{{\bm{k}}}}
\frac{1-f_{{\bm{k}}}}{1-f_{{\bm{q}}}}
\;\;\;\;\;
\gamma_{{\bm{k}}}
=
\int  d{\bm{q}}\;
\gamma_{{\bm{k}}{\bm{q}}}
\frac{1-f_{{\bm{q}}}}{1-f_{{\bm{k}}}}
\label{Gamma_and_M}
\end{equation}
and adopting an Einstein convention with implied integration over the
momentum ${\bm{q}}$, but not ${\bm{k}}$, we may write the Boltzmann
equation in the form
\begin{equation}
\left[ \partial_t +(e{\bm{E}}/\hbar)\cdot\partial_{{\bm{k}}} \right]f_{{\bm{k}}}
=
- \gamma_{{\bm{k}}} [1- \mathsf{M}]_{{\bm{k}}{\bm{q}}} f_{{\bm{q}}}
(1-f_{{\bm{q}}}).
\label{Boltzmann2}
\end{equation}
Let us consider an initial thermal distribution of electrons and
auxiliary modes at the same temperature.
The deviation in the electron distribution from its initial thermal
distribution, $f^0_{\bm{k}}$, in response to an electric field is given by a solution
of the linearised equation
\begin{equation}
(e{\bm{E}}/\hbar)\cdot\partial_{{\bm{k}}}
\left[ f^0_{{\bm{k}}} + \delta f_{{\bm{k}}}  \right]
=
- \gamma_{{\bm{k}}}
[1- \mathsf{M}]_{{\bm{k}}{\bm{q}}}
\delta f_{{\bm{q}}},
\label{linearize}
\end{equation}
where an Einstein convention has again been adopted.
There are a couple of steps required in deriving this
equation. Firstly, we have used the fact that the scattering
integral is zero when the electrons and magnons are in thermal
distributions at the same temperature. One must also allow for the
dependence of  $[1- \mathsf{M}]_{{\bm{k}}{\bm{q}}} $ upon $f_{\bm{q}}$ in obtaining
the first functional derivative of the scattering integral.

A formal solution to the linearised Boltzmann equation (\ref{linearize}) is readily obtained.
Expanding to linear order in the electrical field we find
\begin{equation}
\delta f_{{\bm{k}}}
= [1- \mathsf{M}]^{-1}_{{\bm{k}}{\bm{q}}}
\frac{1}{\gamma_{{\bm{q}}}} {\bm{E}}\cdot \partial_{{\bm{q}}}
f_{{\bm{q}}}.
\label{df2}
\end{equation}
This result may be integrated to obtain the current that flows in
response to the application of the electric field;
\begin{eqnarray}
{\bm{j}}
&=&
\int
\frac{d {\bm{k}}}{(2 \pi)^d}
{\bm{k}} \delta f_{\bm{k}}
\nonumber\\
&=&
\int
\frac{d {\bm{k}}}{(2 \pi)^3}
\frac{d {\bm{q}}}{(2 \pi)^3}
{\bm{k}}
[1- \mathsf{M}]^{-1}_{{\bm{k}}{\bm{q}}}
\frac{1}{\gamma_{{\bm{q}}}} {\bm{E}}\cdot \partial_{{\bm{q}}}
f_{{\bm{q}}}
,
\end{eqnarray}
where an explicit integral over ${\bm{k}}$ has been restored.
In the next section, we will turn to a consideration of the non-linear
response of the electron-magnon system using a very similar Boltzmann
transport analysis. Before this, we identify a number of different
time-scales of relevance to our problem and calculate them in the case
of magnon scattering.

\subsection{A Compendium of Relaxation Rates}
The electronic scattering integral is given by the right-hand side of
Eq.~(\ref{Electron_Boltzmann1}) or Eq.(\ref{linearize}),
\begin{eqnarray}
\left(
\frac{\partial f_{\bm{q}}}{\partial t}
\right)_{\textrm{Scatt}}
&=&
-
\int \frac{d {\bm{p}}}{(2 \pi)^3}
\left[
\begin{array}{l}
\gamma_{{\bm{q}}{\bm{p}}} f_{\bm{q}} (1- f_{\bm{p}})
\\
\;\;\;\;\;
-
\gamma_{{\bm{p}}{\bm{q}}} f_{\bm{p}}(1-f_{\bm{q}})
\end{array}
\right]
\nonumber\\
&=&
-
\gamma_{\bm{q}} [1-\mathsf{M}]_{{\bm{q}}{\bm{p}}} f_{\bm{p}} (1-f_{\bm{p}})
\nonumber\\
&=&
-
\gamma_{\bm{q}} [1-\mathsf{M}]_{{\bm{q}}{\bm{p}}} \delta f_{\bm{p}} ,
\end{eqnarray}
where
$\gamma_{\bm{q}}$
and
$
\mathsf{M}_{{\bm{q}}{\bm{p}}}
$
are given by Eq.~(\ref{Gamma_and_M}).
Integration over ${\bm{p}}$ has been suppressed in the final two
expressions, which are drawn from
Eqs.~(\ref{Boltzmann2}) and (\ref{linearize}), respectively.

We may identify several different time-scales from this scattering
integral that will appear in our later study of the
non-equilibrium response;
\begin{eqnarray}
\frac{1}{\tau_{\bm{q}}}
&=&
\gamma_{\bm{q}}
=
\int \frac{d {\bm{p}}}{(2 \pi)^3}
\gamma_{{\bm{q}}{\bm{p}}}
\frac{1-f_{\bm{p}}}{1-f_{\bm{q}}}
,
\nonumber\\
\frac{1}{\tau^{\textrm{tr}}_{\bm{q}}}
&=&
\gamma^{\textrm{tr}}_{\bm{q}}
=
\int \frac{d {\bm{p}}}{(2 \pi)^3}
\gamma_{{\bm{q}}{\bm{p}}}
\frac{1-f_{\bm{p}}}{1-f_{\bm{q}}}
\left[
1
-
\frac{{\bm{q}}.{\bm{p}}}{q^2}
\frac{\gamma_{\bm{q}}^{\textrm{tr}}}{\gamma_{\bm{p}}^{\textrm{tr}}}
\right]
,
\nonumber\\
\frac{d {\cal E}}{dt}
&=&
\frac{1}{2}
\int \frac{d {\bm{p}}}{(2 \pi)^3}\frac{d {\bm{q}}}{(2 \pi)^3}
(\epsilon_{\bm{q}}-\epsilon_{\bm{p}})
\left[
\begin{array}{l}
\gamma_{{\bm{q}}{\bm{p}}}
f^0_{\bm{q}}(1-f^0_{\bm{p}})
\\
\;\;\;
-
\gamma_{{\bm{p}}{\bm{q}}}
f^0_{\bm{p}}(1-f^0_{\bm{q}})
\end{array}
\right]
.
\nonumber\\
\label{timescales}
\end{eqnarray}
The first of these scattering rates is simply the inverse time between
collisions. The second is the transport scattering rate. This has the
usual additional geometrical factor arising since large angle
scattering has more effect upon transport than small angle
scattering\cite{footnote3}. The ratio
$\gamma_{\bm{q}}^{\textrm{tr}}/ \gamma_{\bm{p}}^{\textrm{tr}}$
is conventionally set to $1$ since we are interested in
the scattering of fermions near to the Fermi surface. The final
expression is the rate of flow of energy from auxiliary modes at a
temperature $T$ (at which $\gamma_{{\bm{p}}{\bm{q}}}$ is evaluated) to
electrons at temperature $T_0$ (indicated by the superscript $0$ on the
electron distribution functions).

In the high-temperature limit, these relaxation rates have the
following temperature dependence within the Moriya-Hertz-Millis~\cite{Moriya, Hertz76, Millis} theory in  $d$ dimensions:
\begin{eqnarray}
1/\tau
&\propto&
T^{(d+z-3)/z},
\nonumber\\
1/\tau_{\textrm{tr}}
&\propto&
T^{(d+z-1)/z},
\nonumber\\
d{\cal E}/dt
&\propto&
T^{(d+3z-3)/z}.
\label{rates}
\end{eqnarray}
Details of how to get these results from Eqs.(\ref{timescales}) are
given in Appendix A. After these preliminaries, we are now in a
position to adapt our Boltzmann equation to describe the
out-of-equilibrium behaviour of our system.

\section{Non-equilibrium Response}

In this section we will turn the machinery of Boltzmann transport to
the question of non-equilibrium behaviour in the itinerant critical
ferromagnet. As discussed earlier, for a system to have an
out-of-equilibrium steady state under the application of an electric
field, it must be coupled to a heat sink that can dissipate the excess
energy generated by Joule heating. We must pay careful attention to
the various thermal couplings. The nature of these bears repetition at
this juncture.

As described in the introduction, we consider a
long, narrow sample in which the excess Joule heat is carried away by a
transverse heat current to a heat sink at the edge. Heat entering the
electrons {\it via} Joule heating passes to the magnons and then back to the electrons
 {\it via} mutual scattering and leaves the electrons {\it via} coupling to a
heat sink at the boundary. The magnons themselves interact both with
the electrons and with one another. Electron-electron scattering is
neglected in our analysis--- it is higher order in temperature and
hence field --- as is
coupling of magnons directly to the heat sink.

Our analysis is divided into three parts. We begin by writing down
the Boltzmann equations for the electron-magnon system. These
equations embody the various thermal couplings and interactions in our
system. The only subtlety enters through the form of the magnons'
Boltzmann equation: the overdamped nature of the magnon
excitations leads to a slightly more complicated equation than the
comparable case of phonon scattering. In fact, the details of the
magnon Boltzmann equation will not have a huge effect upon our main
result. Next, we will present formal solutions for the electron and
magnon distribution functions. Our main results follow from
consideration of these solutions in the limit where magnon-magnon
scattering leads to a thermal distribution of magnons with temperature
determined by the electric field. We will end with an argument why
correction to this thermal distribution of magnons do not change the
scaling of our results.

\subsection{The Boltzmann Equation}
{\it The electron Boltzmann equation} is given by a minimal modification of
the thermal equilibrium Boltzmann equation (\ref{Electron_Boltzmann1}):
\begin{eqnarray}
& &
\left[
\partial_t
+(e {\bm{E}}/\hbar)\cdot\partial_{{\bm{k}}}
+{\bm{v}}_{\bm{k}} \cdot \nabla
\right]
f_{{\bm{k}}}
\nonumber\\
&=&
\mathsf{I}^{em}_{\bm{k}}[f,n]
+\hbox{Scattering to heat sink}
\nonumber\\
&=&
-\int \frac{d{\bm{q}}}{(2 \pi)^3} \;
\left[
\begin{array}{l}
\gamma_{{\bm{k}}{\bm{q}}} f_{\bm{k}} \left( 1 - f_{\bm{q}} \right)
\;\;\;\;
\\
\;\;\;\;\;
-
\gamma_{\bm{q}\bm{k}} f_{\bm{q}} \left( 1 - f_{\bm{k}} \right)
\end{array}
\right]
+\hbox{heat sink}
\nonumber\\
\label{Electron_Boltzmann2}
\end{eqnarray}
where $\mathsf{I}^{em}_{\bm{k}}[f,n]$ indicates the scattering integral
for electrons of momentum ${\bm{k}}$ scattering from magnons.
The gradient term ${\bm{v}}_{\bm{k}}. \nabla$ has been added to allow for
 the possibility of transverse heat flow.
The electron-magnon scattering matrices now take a slightly modified
form compared with that in thermal equilibrium (\ref{MagnonGamma}).
Since the magnons are overdamped and as a result do not have a definite
relationship between their energy and momentum, the magnon
distribution is a function of both energy and momentum. Taking this
into account, the scattering matrices take the form
\begin{equation}
\gamma_{{\bm{p}}{\bm{q}}}
=
|g_{{\bm{q}}-{\bm{p}}}|^2
\left[
\begin{array}{l}
\left[
1+n_{{\bm{p}}-{\bm{q}}}(\epsilon_{\bm{p}}-\epsilon_{\bm{q}}) \right]
\rho({\bm{p}}-{\bm{q}}, \epsilon_{\bm{p}}-\epsilon_{\bm{q}})
\\
\;\;\;\;\;\;\;\;
+n_{{\bm{q}}-{\bm{p}}}(\epsilon_{\bm{q}}-\epsilon_{\bm{p}})
\rho({\bm{q}}-{\bm{p}}, \epsilon_{\bm{q}}-\epsilon_{\bm{p}})
\end{array}
\right].
\label{MagnonGamma2}
\end{equation}
The linearised expansion about the zero-field, base-temperature distribution $f^0_{\bm{q}}$ takes the form
\begin{eqnarray}
& &
(e{\bm{E}}/\hbar)\cdot\partial_{{\bm{k}}}
\left[ f^0_{{\bm{k}}} + \delta f_{{\bm{k}}}  \right]
\nonumber\\
&=&
-\int \frac{d{\bm{q}}}{(2 \pi)^3} \;
\left[
\begin{array}{l}
\gamma_{{\bm{k}}{\bm{q}}} f_{\bm{k}} \left( 1 - f_{\bm{q}} \right)
\;\;\;\;
\\
\;\;\;\;\;
-
\gamma_{\bm{q}\bm{k}} f_{\bm{q}} \left( 1 - f_{\bm{k}} \right)
\end{array}
\right]
\nonumber\\
& &
- \gamma_{{\bm{k}}}
[1- \mathsf{M}]_{{\bm{k}}{\bm{q}}}
\delta f_{{\bm{q}}}
+\hbox{heat sink}
\label{Linearized_OOE}
\end{eqnarray}
where $\gamma_{\bm{k}}$ and $\mathsf{M}_{\bm{k}\bm{q}}$ take a slightly modified form
out of equilibrium given by
\begin{eqnarray}
\gamma_{\bm{k}}
&=&
\int \frac{d {\bm{q}} }{(2 \pi)^3}
\left[
\gamma_{{\bm{k}}{\bm{q}}} \left( 1-f^0_{\bm{q}} \right)
+
\gamma_{{\bm{q}}{\bm{k}}} f^0_{{\bm{q}}}
\right]
\nonumber\\
\gamma_{\bm{k}} \mathsf{M}_{{\bm{k}}{\bm{q}}}
&=&
\gamma_{{\bm{k}}{\bm{q}}} f^0_{\bm{k}}
+
\gamma_{{\bm{q}}{\bm{k}}} \left( 1-f^0_{\bm{k}} \right).
\label{GeneralMGamma}
\end{eqnarray}
These reduce to our previous expressions for $\gamma_{\bm{k}}$ and $\mathsf{M}_{\bm{k}\bm{q}}$
in thermal equilibrium (\ref{Gamma_and_M}). The simplified forms given by (\ref{Gamma_and_M})
can be obtained by making use of the detailed balance condition, which is not satisfied out of equilibrium. A couple of points are worth making about Eq.(\ref{Linearized_OOE}). Firstly, there is a zeroth order term on the right-hand side. This term is not present in thermal equilibrium (it is zero upon applying the detailed balance condition). This term has a different symmetry in momentum space than the first order term in $\delta f$ and we will use this in our analysis shortly.

The added complication
 due to the magnons being overdamped is compounded when we come to
 write down the magnon Boltzmann equation in a moment. Luckily, this
 does not affect the bulk of our calculation. We will use the
 formal notation $\mathsf{I}^{em}_{\bm{k}}[f,n]$ through as much of our
 analysis as possible in order to keep algebra to a minimum. When we
 eventually substitute the particular form of the scattering integrals
 near to the end of the calculation, we will find that most of the
 integration of these scattering integrals carries over directly from
 the thermal equilibrium calculation.

\begin{widetext}
\noindent
{\it The Magnon Boltzmann Equation} takes the form
%
\begin{eqnarray}
\partial_t n_{\bm{k}}(\epsilon)
&=&
\mathsf{I}^{me}_{\bm{k}}[f,n]
+\mathsf{I}^{mm}_{\bm{k}}[n]
\nonumber\\
&=&
\int
\frac{d{\bm{p}}}{(2 \pi)^3}
\frac{d{\bm{q}}}{(2 \pi)^3}
|g_{{\bm{k}}}|^2
\left[
\begin{array}{l}
-n_{\bm{k}}(\epsilon )
f_{{\bm{p}}} (1-f_{{\bm{q}}})
\\
\;\;\;
+[1+n_{\bm{k}}(\epsilon) ]
(1-f_{{\bm{p}}})f_{{\bm{q}}}
\end{array}
\right]
\delta (\epsilon+\epsilon_{\bm{p}}-\epsilon_{\bm{q}})
\delta ({\bm{p}}-{\bm{q}} +{\bm{k}})
\nonumber\\
& &
+
\lambda
\int
d \epsilon_1 d \epsilon_2 d \epsilon_3
\frac{d{\bm{p}}_1}{(2 \pi)^3}
\frac{d{\bm{p}}_2}{(2 \pi)^3}
\frac{d{\bm{p}}_3}{(2 \pi)^3}
\rho (\epsilon_1, {\bm{p}}_1)
\rho (\epsilon_2, {\bm{p}}_2)
\rho (\epsilon_3, {\bm{p}}_3)
\nonumber\\
& &
\;\;\;\;\;\;\;\;\;\;\;\;\;\;\;\;\;
\times
\left[
\begin{array}{l}
-n_{\bm{k}}(\epsilon) n_{{\bm{p}}_1}(\epsilon_1)
[1+n_{{\bm{p}}_2}(\epsilon_2 ) ][1+n_{{\bm{p}}_3}(\epsilon_3) ]
\\
\;\;\;\;\;
 +[ 1+n_{{\bm{k}}}(\epsilon) ]
[1+ n_{{\bm{p}}_1}(\epsilon_1) ]
n_{{\bm{p}}_2}(\epsilon_2 ) n_{{\bm{p}}_3}(\epsilon_3)
\end{array}
\right]
\delta ( {\bm{k}}+{\bm{p}}_1-{\bm{p}}_2-{\bm{p}}_3)
\delta ( \epsilon + \epsilon_1- \epsilon_2 - \epsilon_3).
\nonumber\\
\label{Magnon_Boltzmann}
\end{eqnarray}
\end{widetext}
The easiest way to see the origins of the various terms in this equation is to momentarily treat the magnons as if they had a definite relationship between energy and momentum. In this case, the magnon spectral function becomes a delta-function and the scattering integrals reduce to the same form as those for electron-phonon scattering. As for the electron Boltzmann equation, we will carry out
as much of our analysis as possible using the formal expressions
$\mathsf{I}^{me}_{\bm{k}}[f,n]$ and
$\mathsf{I}^{mm}_{\bm{k}}[n]$ for the magnon-electron and magnon-magnon
scattering integrals.

\subsection{Solving the Boltzmann Equations}
Our analysis of the Boltzmann equations
(\ref{Electron_Boltzmann2}) and (\ref{Magnon_Boltzmann}) derived above
proceeds as follows:
We begin by dividing the electron distribution function into two
parts--- a spherically symmetric part and a non-symmetric part. The magnon
Boltzmann equation is divided similarly.
After this division, the resulting Boltzmann equations have simple
interpretations. The equation for the
symmetric part of the distribution function describes the balance
between the transverse heat flow and the flow of energy out of the
magnons into the symmetric part of the electron distribution. The
equation for the remaining part describes a balance between the flow
of energy from the electrons into the magnons and the Joule heating
rate. In order to obtain useful results from these equations, we will
go to a limit where the magnon distribution is assumed to be
thermalised at some temperature $T_{\textrm{eff}}(E)$. The final step in our
analysis will be to show that corrections to the thermal distribution
of magnons do not change the way in which the current response scales
with field.

\vspace{0.1in}
\noindent
{\it Expanding the Boltzmann Equation}:

\noindent
The electron distribution function is divided into its symmetric part
$f^0$ (assumed to be a Fermi distribution at a low base temperature
that varies slowly across the sample) and the
remainder $\delta f$. With this notation and after expanding to linear
order in $\delta f$, the Boltzmann equations may be written in the
form
\begin{eqnarray}
{\bm{v}}_{\bm{q}} \cdot \nabla f^0_{\bm{q}}
&=&
\left[
\mathsf{I}^{em}_{\bm{q}}[f^0,n]
\right]^S,
\label{one}
\\
{\bm{E}} \cdot \partial_{\bm{q}}
\left( f^0_{\bm{q}} + \delta f_{\bm{q}} \right)
&=&
\left[
\mathsf{I}^{em}_{\bm{q}}[f^0,n]
\right]^A
+
\frac{ \delta \mathsf{I}^{em}_{\bm{q}}}
{\delta f_{\bm{k}}} \delta f_{\bm{k}},
\label{two}
\\
0
&=&
\mathsf{I}^{me}_{{\bm{k}},\epsilon}[f^0,n]
+
\frac{ \delta \mathsf{I}^{me}_{{\bm{k}},\epsilon}}
{\delta f_{\bm{q}}} \delta f_{\bm{q}}
+
\mathsf{I}^{mm}_{{\bm{k}},\epsilon}[n].
\nonumber\\
\label{three}
\end{eqnarray}
We have adopted an Einstein convention where terms like
$(\delta \mathsf{I}^{em}_{\bm{q}}/
\delta f_{\bm{k}}) \delta f_{\bm{k}}$
are implicitly integrated over ${\bm{k}}$. The superscripts $S$ and $A$
refer to symmetric and non-symmetric parts of the scattering integrals
in ${\bm{q}}$. We have allowed for a transverse gradient in $f^0$ which
supports a transverse heat flow.

One might question how, given the fact that we are interested in the
non-linear response, we can use a linear analysis in $\delta f$.
In a linear-response, relaxation-time
approximation, the Fermi surface is effectively shifted a distance
$\tau_{\textrm{tr}} e E/ \hbar$ in momentum space. Provided this is much less than
the Fermi wavevector ($\tau_{\textrm{tr}} e E/ \hbar\ll k_{\textrm{F}}$) a linear response
analysis may be applied. In the present case, it turns out that the
transport relaxation time, $\tau_{\textrm{tr}}$, self-consistently becomes a
power of $E$ so that the resultant current is non-linear in $E$.

\vspace{0.1in}
\noindent
{\it Heat flows}

\noindent
The physical content of the equations (\ref{one}), (\ref{two}) and
(\ref{three}) is most readily appreciated by considering the energy
transfers that they represent. In the case of (\ref{one}) and (\ref{two})
we multiply by the electron energy $\epsilon_{\bm{q}}$ and integrate
over ${\bm{q}}$. In the case of (\ref{three}), we multiply by the magnon
energy $\epsilon$ and the spectral density $\rho({\bm{k}}, \epsilon)$
and integrate over ${\bm{k}}$ and $\epsilon$. After doing this,
Eqs.(\ref{one}-\ref{three}) reduce to
\begin{widetext}
\begin{eqnarray}
0&=&
\int
\frac{d {\bm{q}}}{(2 \pi)^3}
\epsilon_{\bm{q}}
\left[
{\bm{v}}_{\bm{q}} \cdot \nabla f^0_{\bm{q}}
-
\left[
\mathsf{I}^{em}_{\bm{q}}[f^0,n]
\right]^S
\right],
\label{energy1}
\\
0
&=&
\int
\frac{d {\bm{q}}}{(2 \pi)^3}
\epsilon_{\bm{q}}
\left[
{\bm{E}} \cdot \partial_{\bm{q}}
\left( f^0_{\bm{q}} + \delta f_{\bm{q}} \right)
-
\left[
\mathsf{I}^{em}_{\bm{q}}[f^0,n]
\right]^A
-
\frac{ \delta \mathsf{I}^{em}_{\bm{q}}}
{\delta f_{\bm{k}}}[f^0,n] \delta f_{\bm{k}}
\right],
\label{energy2}
\\
0
&=&
\int
\frac{d {\bm{k}}}{(2 \pi)^3}d \epsilon
\rho(\epsilon, {\bm{k}})
\epsilon
\left[
\mathsf{I}^{me}_{{\bm{k}},\epsilon}[f^0,n]
+
\frac{ \delta \mathsf{I}^{me}_{{\bm{k}},\epsilon}}
{\delta f_{\bm{q}}}[f^0,n] \delta f_{\bm{q}}
+
\mathsf{I}^{mm}_{{\bm{k}},\epsilon}[n]
\right].
\label{energy3}
\end{eqnarray}
\end{widetext}
Eq.(\ref{energy1}) may be interpreted as a balance between the
transverse heat flow --- described by the first term on the
right-hand side --- and the energy flowing into the symmetrical part of
the electron distribution --- described by the second term on the right
hand side. The flow of heat into the heat sink has been
treated as a boundary condition in writing down this equation. Solving
this equation leads to the explicit limit on the sample width
discussed in the introduction and worked out in detail in
Ref.\cite{Tremblay79}. We will not concentrate upon it further here.

\vspace{0.1in}
\noindent
Eq.(\ref{energy2}) can be interpreted as a balance between Joule
heating --- described by the first term on the right-hand-side --- and the
rate at which energy flows from the non-symmetrical part of the
electron distribution into the magnons --- described by the second and
third terms. To see this requires a little manipulation. The first
term may be written explicitly as Joule heating after integrating by
parts with respect to ${\bm{q}}$.

\vspace{0.1in}
\noindent
Eq.(\ref{energy3}) corresponds to a balance between the rate at which
energy flows into the magnons from the symmetrical and non-symmetrical
parts of the electron distribution --- described by the first and second
terms respectively. The net flow of energy into
the magnons from the electrons is zero in a steady state, if we neglect heat flow directly from the magnons to the heat sink. The latter process is ignored since it is much slower than magnon-electron scattering.
The third term is identically zero, since
magnon-magnon scattering conserves energy. This fact is extremely
useful. By considering this integrated equation, one can avoid having
to deal explicitly with the magnon-magnon scattering integral.

We can transform this final equation into a more useful form by using
the fact that electron-magnon scattering is energy conserving. This
implies that
\begin{widetext}
\begin{eqnarray*}
\int
\frac{d {\bm{q}}}{(2 \pi)^3}
\epsilon_{\bm{q}}
\mathsf{I}^{em}_{\bm{q}}[f^0,n]
&=&
\int
\frac{d {\bm{k}}}{(2 \pi)^3}d \epsilon
\;\; \epsilon \rho(\epsilon, {\bm{k}})
\mathsf{I}^{me}_{{\bm{k}},\epsilon}[f^0,n]
\\
\int
\frac{d {\bm{q}}}{(2 \pi)^3}
\epsilon_{\bm{q}}
\frac{
 \delta \mathsf{I}^{em}_{\bm{q}}[f^0,n]}
{\delta f_{\bm{k}}} \delta f_{\bm{k}}
&=&
\int
\frac{d {\bm{k}}}{(2 \pi)^3}d \epsilon
\;\;
\epsilon
\rho(\epsilon, {\bm{k}})
\frac{ \delta \mathsf{I}^{me}_{{\bm{k}},\epsilon}[f^0,n]}
{\delta f_{\bm{q}}} \delta f_{\bm{q}}
\end{eqnarray*}
\end{widetext}
{\it i.e} the energy entering the electrons from the magnons is equal
to the energy entering the magnons from the electrons. Using these
results reduces Eq.(\ref{energy3}) to the form
\begin{equation}
0
=
\int
\frac{d {\bm{q}}}{(2 \pi)^3}
\epsilon_{\bm{q}}
\left[
\mathsf{I}^{em}_{\bm{q}}[f^0,n]
+
\frac{ \delta \mathsf{I}^{em}_{\bm{q}}[f^0,n]}
{\delta f_{\bm{k}}} \delta f_{\bm{k}}
\right].
\label{energy3_1}
\end{equation}
To make further progress, we must solve the Boltzmann equations
explicitly. We will do this to leading order through an approximation
that the magnon distribution is thermal. We will then argue that
corrections to this do not alter the scaling.

\vspace{0.1in}
\noindent
{\it Thermal Magnon Approximation}

\noindent
Our leading approximation is to assume a thermal distribution of
magnons,
$n_{\bm{k}}(\epsilon)= n^0(\epsilon)=(\textrm{e}^{\epsilon/T_{\textrm{eff}}}-1)^{-1}$,
where the effective temperature is to be determined shortly. In this
case, the linearised Boltzmann equations (\ref{one}), (\ref{two}) and (\ref{three}) reduce to
\begin{eqnarray}
{\bm{v}}_{\bm{q}} \cdot \nabla f^0_{\bm{q}}
=
\mathsf{I}^{em}_{\bm{q}}[f^0,n^0],
\label{A}
\\
{\bm{E}} \cdot \partial_{\bm{q}}
\left( f^0_{\bm{q}} + \delta f_{\bm{q}} \right)
=
\frac{ \delta \mathsf{I}^{em}_{\bm{q}}[f^0,n^0]}
{\delta f_{\bm{k}}} \delta f_{\bm{k}},
\label{B}
\\
0
=
\mathsf{I}^{me}_{{\bm{k}},\epsilon}[f^0,n^0]
+
\frac{ \delta \mathsf{I}^{me}_{{\bm{k}},\epsilon}[f^0,n^0]}
{\delta f_{\bm{q}}} \delta f_{\bm{q}}.
\label{C}
\end{eqnarray}
We have used the fact that the magnon-magnon scattering is identically
zero for a thermal distribution of magnons. The second equation may be
formally solved for $\delta f$ to obtain,
\begin{equation}
\delta f_{\bm{q}}
=
\left[
1-
\left(\frac{ \delta \mathsf{I}^{em}}
{\delta f}\right)^{-1} {\bm{E}}. \partial_{\bm{q}}
\right]^{-1}
\left(\frac{ \delta \mathsf{I}^{em}}
{\delta f}\right)^{-1} {\bm{E}}. \partial_{\bm{q}}
f^0,
\end{equation}
where we have suppressed momentum labels and integrals over momentum
for brevity. Expressions such as
$\left(\delta \mathsf{I}^{em}/
\delta f\right)^{-1}$
are to be understood as matrix inverses with appropriate integrations
over momentum in their products. In order to determine the effective
temperature, we substitute this solution for $\delta f$ into the
energy-integrated form of Eq.(\ref{energy3}) or (\ref{B})
given by Eq.(\ref{energy3_1})
and expand to leading order in ${\bm{E}}$. The result of this
substitution is
\begin{equation}
0
=
\int
\frac{d {\bm{q}}}{(2 \pi)^3}
\epsilon_{\bm{q}}
\left[
\mathsf{I}^{em}_{\bm{q}}
+
{\bm{E}} \cdot \partial_{\bm{q}}
\left[
\left(
\frac{ \delta \mathsf{I}^{em}}
{\delta f}\right)^{-1}
{\bm{E}} \cdot \partial_{\bm{q}}
 \delta f
\right]
\right].
\end{equation}
Integrating the second term by parts reduces it to the form
\begin{equation}
0
=
\int
\frac{d {\bm{q}}}{(2 \pi)^3}
\epsilon_{\bm{q}}
\left[
\mathsf{I}^{em}_{\bm{q}}
-
\frac{1}{3}
E^2
{\bm{v}}_{\bm{q}}
\cdot
\left(
\frac{ \delta \mathsf{I}^{em}}
{\delta f}\right)^{-1}
\partial_{\bm{q}}
 \delta f
\right].
\label{general_balance}
\end{equation}
The second term is now explicitly the leading order contribution to
the Joule heating rate. This is balanced against the first term which
describes the decay of energy from a thermal distribution of magnons
at temperature $T_{\textrm{eff}}(E)$. Since both the electron and magnon
distributions involved in the expressions are thermal distributions ---
at $T=0$ and $T=T_{\textrm{eff}}(E)$ respectively --- we may evaluate
Eq.(\ref{general_balance})
using the results of Section 3. Writing Eq.(\ref{general_balance})
in terms of the scattering matrices of Section 3, it can be reduced to
\begin{eqnarray}
& &
\int
\frac{ d{\bm{q}}}{(2 \pi)^3}
\epsilon_{\bm{q}}
\gamma_{\bm{q}} [1-\mathsf{M}]_{{\bm{q}}{\bm{p}}} f_{\bm{p}}(T_0)
\left( 1-f_{\bm{p}}(T_0) \right),
\nonumber\\
&=&
\int
\frac{d{\bm{q}}}{(2 \pi)^3}
\epsilon_{\bm{q}}
\gamma_{\bm{q}} [1-M]_{{\bm{q}}{\bm{p}}} \delta f_{\bm{p}}.
\label{ThermalBalance1}
\end{eqnarray}
Substituting for $\delta f$ to
leading order in ${\bm{E}}$ from Eq.~(\ref{B}) into
Eq.~(\ref{ThermalBalance1}), we obtain
\begin{eqnarray}
& &
\underbrace{
\int \frac{d {\bm{q}}}{(2 \pi)^3}
\epsilon_{\bm{q}}.\partial_{\bm{q}}
\left[
[1-\mathsf{M}]_{{\bm{q}}{\bm{p}}}^{-1} \gamma^{-1}_{\bm{p}}
{\bm{E}}.\partial_{\bm{q}} f_{\bm{q}}(T_0)
\right]
}_{
\hbox{Joule Heating}}
\nonumber\\
&=&
\underbrace{
\int \frac{d {\bm{q}}}{(2 \pi)^3}
\epsilon_{\bm{q}}
\gamma_{\bm{q}}
[1-\mathsf{M}]_{{\bm{q}}{\bm{p}}}
f_{\bm{p}}(T_0)
\left[1-f_{\bm{p}}(T_0)
\right]
}_{
\hbox{Energy decay from } T_{\textrm{eff}} \hbox {to } T_0}
\label{ThermalBalance2}
\end{eqnarray}
This equation may be written in the form $d{\cal E}/dt \propto E^2 \tau_{\textrm{tr}} $
 as before in Eq,(\ref{EnergyBalance}). Since the magnon distribution is thermal at temperature $T_{\textrm{eff}}$, using the temperature scaling of the various relaxation rates given in Eq.(\ref{rates}), we find
$$
T_{\textrm{eff}}
\propto
E^{z/(d+2z-2)}
$$
in the high-field/temperature limit and $d$ dimensions,
implying a non-linear current
$$
j
\propto
E^{(z-1)/(d+2z-2)}
$$
as suggested in Section 1.

\vspace{0.1in}
\noindent
{\it Corrections to Thermal Magnon Approximation}

\noindent
Corrections to a thermal distribution of magnons may be rather large,
since in the absence of the electric field the magnons are
essentially in a zero-temperature
distribution. We argue, nevertheless, that corrections to the thermal
distribution of magnons considered above do not change the scaling of
current response. The analysis is similar to the
calculation of phonon drag in thermal equilibrium (which similarly
does not change the scaling with temperature).

We expand the magnon distribution to linear order about the effective
thermal distribution; $n=n^0 + \delta n$. Substituting into the
Boltzmann Eqs.(\ref{one},\ref{two}) and (\ref{three}), we obtain
\begin{eqnarray}
{\bm{v}}_{\bm{q}} \cdot \nabla f^0_{\bm{q}}
&=&
\mathsf{I}^{em}_{\bm{q}}[f^0,n^0]
+
\frac{\delta \mathsf{I}^{em}_{\bm{q}}[f^0,n^0]}
{\delta n_{{\bm{k}},\epsilon}}
\delta n^S_{{\bm{k}},\epsilon},
\label{AA}
\\
{\bm{E}} \cdot \partial_{\bm{q}}
\left( f^0_{\bm{q}} + \delta f_{\bm{q}} \right)
&=&
\frac{ \delta \mathsf{I}^{em}_{\bm{q}}[f^0,n^0]}
{\delta f_{\bm{k}}} \delta f_{\bm{k}}
+
\frac{ \delta \mathsf{I}^{em}_{\bm{q}}[f^0,n^0]}
{\delta n_{{\bm{k}},\epsilon}}
\delta n^A_{{\bm{k}},\epsilon},
\label{BB}
\nonumber\\
\\
0
&=&
\mathsf{I}^{me}_{{\bm{k}},\epsilon}[f^0,n^0]
+
\frac{ \delta \mathsf{I}^{me}_{{\bm{k}},\epsilon}[f^0,n^0]}
{\delta f_{\bm{q}}} \delta f_{\bm{q}}
\\
& &
+
\frac{ \delta \mathsf{I}^{me}_{{\bm{k}},\epsilon}[f^0,n^0]}
{\delta n_{{\bm{q}},\xi}}
\delta n^S_{{\bm{q}},\xi}
+
\frac{ \delta \mathsf{I}^{mm}_{{\bm{k}},\epsilon}[n^0]}
{\delta n_{{\bm{q}},\xi}}
\delta n_{{\bm{q}},\xi}.
\label{CC}
\nonumber\\
\end{eqnarray}
In Eqs.(\ref{AA}) and (\ref{BB}), $\delta n$ has been divided into
symmetric and non-symmetric parts $\delta n^S$ and $\delta n^A$. These
contribute to the equations for the symmetric and non-symmetric parts
of the electron distribution respectively. Eq.(\ref{CC}) can be solved
formally for $\delta n$ with the result
\begin{eqnarray}
\delta n
&=&
-
\underbrace{
\left(
\frac{\delta \mathsf{I}^{me}}{\delta n}
+
\frac{\delta \mathsf{I}^{mm}}{\delta n}
\right)^{-1}
\mathsf{I}^{me}}_{\delta n^S}
\\
& &
-
\underbrace{
\left(
\frac{\delta \mathsf{I}^{me}}{\delta n}
+
\frac{\delta \mathsf{I}^{mm}}{\delta n}
\right)^{-1}
\frac{ \delta \mathsf{I}^{me}}{\delta f}
\delta f}_{\delta n^A}
\end{eqnarray}
We have identified the spherically symmetric and non-symmetric parts
of $\delta n$. Substituting  this back into Eq.(\ref{AA}) and
(\ref{BB}) one obtains
%
\begin{eqnarray}
{\bm{v}}_{\bm{q}} \cdot \nabla f^0_{\bm{q}}
&=&
\mathsf{I}^{em}_{\bm{q}}[f^0,n^0]
\nonumber\\
& &
-
\frac{
\delta \mathsf{I}^{em}_{\bm{q}}
}{\delta n}
\left(
\frac{\delta \mathsf{I}^{me}}{\delta n}
+
\frac{\delta \mathsf{I}^{mm}}{\delta n}
\right)^{-1}
\mathsf{I}^{me}
\label{AAA}
\\
{\bm{E}} \cdot \partial_{\bm{q}}
\left( f^0_{\bm{q}} + \delta f_{\bm{q}} \right)
&=&
\frac{ \delta \mathsf{I}^{em}_{\bm{q}}[f^0,n^0]}
{\delta f_{\bm{k}}} \delta f_{\bm{k}}
\nonumber\\
& &
\underbrace{-
\frac{ \delta \mathsf{I}^{em}_{\bm{q}}}
{\delta n_{{\bm{k}},\epsilon}}
\left(
\frac{\delta \mathsf{I}^{me}}{\delta n}
+
\frac{\delta \mathsf{I}^{mm}}{\delta n}
\right)^{-1}
\frac{\delta \mathsf{I}^{me}}{\delta f}
}_{
\frac{\delta \mathsf{I}^{eme}}{\delta f}}
\delta f
\nonumber\\
\label{BBB}
\end{eqnarray}
%
In the second of these equations, we have adopted the notation of
Lifshitz-Pitaevskii\cite{Lifshitz} identifying this as a term describing a
magnon-mediated electron-electron interaction. The argument that
magnon drag does not affect scaling is completed by showing that
$\delta \mathsf{I}^{eme}/\delta f$ scales with at least as high a
power of $T$ as $\delta \mathsf{I}^{em}/\delta f$. This requires us to
go beyond the generic form of the scattering integrals to use their
explicit expressions for magnon-electron scattering given in the
Boltzmann equations (\ref{Electron_Boltzmann2}) and
(\ref{Magnon_Boltzmann}). We ignore the magnon-magnon scattering (it
is higher order in $T$ ---and hence $E$--- than the magnon-electron
scattering) we may write
\begin{equation}
\frac{\delta \mathsf{I}^{eme}}{\delta f}
=
-
\frac{ \delta \mathsf{I}^{em}_{\bm{q}}}
{\delta n_{{\bm{k}},\epsilon}}
\left(
\frac{\delta \mathsf{I}^{me}}{\delta n}
+
\frac{\delta \mathsf{I}^{mm}}{\delta n}
\right)^{-1}
\frac{\delta \mathsf{I}^{me}}{\delta f}.
\label{Ieme}
\end{equation}
Taking the explicit form of the scattering integrals, the various
functional derivatives may be written as
\begin{eqnarray*}
\frac{ \delta \mathsf{I}^{em}_{\bm{q}}}
{\delta n_{{\bm{k}},\epsilon}}
&=&
-|g|^2 \rho({\bm{k}}, \epsilon)
\left[
\begin{array}{l}
(f_{\bm{q}}- f_{{\bm{k}}-{\bm{q}}} )
\delta( \epsilon -\epsilon_{\bm{q}} + \epsilon_{{\bm{k}}-{\bm{q}}})
\\
\;\;\;\;\;\;\;
+
(f_{\bm{q}}- f_{{\bm{k}}+{\bm{q}}} )
\delta( \epsilon -\epsilon_{{\bm{q}}+{\bm{k}}} + \epsilon_{{\bm{q}}})
\end{array}
\right]
\\
\frac{\delta \mathsf{I}^{me}_{{\bm{k}}, \epsilon}}{\delta n_{{\bm{l}},\nu}}
&=&
|g|^2\delta ({\bm{l}}-{\bm{k}}) \delta (\nu-\epsilon)
\nonumber\\
& &
\times
\int
\frac{d {\bm{p}}}{(2 \pi)^3}
(f_{{\bm{p}}+{\bm{k}}}-f_{\bm{p}} )
\delta( \epsilon + \epsilon_{\bm{p}}-\epsilon_{{\bm{p}}+{\bm{k}}} )
\\
\frac{\delta \mathsf{I}^{me}_{{\bm{k}} \epsilon}}{\delta f_{\bm{l}}}
&=&
|g|^2
\left[
\begin{array}{l}
-(n_{{\bm{k}}\epsilon}+ f_{{\bm{l}}+{\bm{k}}})
\delta( \epsilon + \epsilon_{\bm{l}} - \epsilon_{{\bm{l}}+{\bm{k}}})
\\
\;\;\;\;\;\;\;
+
(1+n_{{\bm{k}}\epsilon}- f_{{\bm{l}}-{\bm{k}}})
\delta( \epsilon + \epsilon_{{\bm{l}}-{\bm{k}}} - \epsilon_{{\bm{l}}})
\end{array}
\right]
\end{eqnarray*}
Substituting these equations back into Eq.(\ref{Ieme}) shows that the
corrections due to magnon drag lead to contributions to the electron
scattering integral that are at least of the same order in temperature
as the direct contribution.

\section{Conclusions and Prospects}
We have considered non-linear transport near to an itinerant electron
quantum critical point. Since the dynamics near to a quantum critical
point are universal, and since steady-state, out-of-equilibrium
distributions are determined by dynamics, we have argued that the
universality present near to an equilibrium quantum critical point may
be reflected in the out-of-equilibrium behaviour.

There are two ways in which a quantum critical itinerant electron
system may be driven out of equilibrium by an electric field. At the
highest fields, the rate of Joule heating overwhelms the rate at which
heat may be transported out of the system by thermal conduction and the
system heats up until the two balance. This mechanism leads to
non-linear response in truly bulk samples. We have considered
non-linear response in a restricted geometry where we anticipate that
conductivity becomes non-linear at a lower field governed by
the rate at which energy can scatter between electrons and magnons.
The resulting conductivity is expected to be independent of sample
size and geometry (provided that certain constraints are satisfied).
At the lowest fields, the response will return to the linear,
thermal equilibrium response.

The existence of the intermediate range of non-linearity requires restrictions
upon the sample width so that thermal conduction can be maintained at a
sufficient rate to transport away heat generated by Joule heating. The
sample width must nevertheless be sufficient that the magnons
demonstrate their bulk behaviour- {\it i.e} it must be larger than the
magnon correlation length. Because of the rather different scaling of
transport and thermal relaxation lengths with temperature (and hence
field), there is a large window of fields within which the type
of non-linearity that we investigate should exist.

What are the prospects for seeing these effects experimentally?
We have described a particular experimental geometry in which the heat
current is transverse to the electrical current. This enabled the
algebra to be readily negotiated. In order to see these effects
experimentally, we suggest a slightly different geometry
\cite{Santiago}. One possibility is the following: take a bow tie shaped
sample with current injected and removed along opposite wings of the
bow tie. A current sent through this sample should demonstrate a
non-linear steady-state of the type that we have described. The
constriction at the centre of the bow tie will have enhanced field and
current densities and will operate in a non-linear regime. The
injection of current along the extended edge of the bow tie will
reduce contact heating; performing the experiment in a pulsed manner
will further mitigate these effects. The large heat capacity of the wings
of the bow tie compared to the constriction will allow a relatively
long pulse time before heat effects become significant; the wings will
effectively act as a low-temperature heat sink. A sketch of this
arrangement is shown in Fig. 1.

\begin{figure}[hbt]
\centerline{\includegraphics[width=3in]{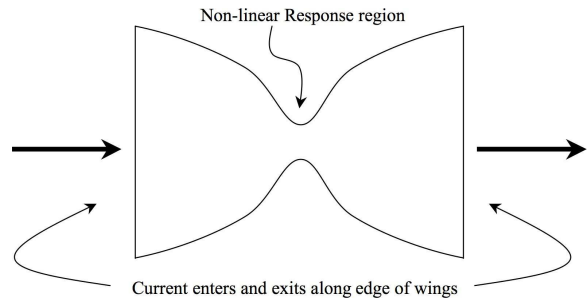}}
\caption{{\bf Schematic diagram of proposed experimental system:} i. Current enters and leaves the bow
tie shaped sample along the edges of the wings, reducing contact heating. ii. Enhanced field and
current density in the constriction leads to non-linear response in this regime. iii. The extended wings act
as a low-temperature heat sink. }
\label{fig:ExpFig}
\end{figure}

It remains to estimate what field strengths give rise to the
non-linear effects that we anticipate. For a typical quantum-critical itinerant
magnet (e.g. $\textrm{Sr}_3\textrm{Ru}_2\textrm{O}_7$, which has
$n=2 \times 10^{27}\,\textrm{m}^{-3}$,
$\sigma= 10^8\, \Omega^{-1}\textrm{m}^{-1}$)
and typical cryogenic temperatures of around $100\, \textrm{mK}$, the electric
field required to observe these non-linear effects is of the order
$$
E= \frac{k_{\textrm{B}} T}{ev_{\textrm{F}} \tau}
\approx 50\, \textrm{V}\textrm{m}^{-1}
$$
(We have used the Drude formula
$\sigma=ne^2 \tau/m$ and $n=4 \pi k_{\textrm{F}}^3/3$)
or a voltage drop of $0.25\,\textrm{V}$ over a typical sample length of $5\,\textrm{mm}$.

In conclusion, the universal power-law scaling of
conductivity near to an itinerant, magnetic, quantum critical point is
reflected in a universal power law scaling with electric field in the
non-linear conductivity regime. This provides a new way to investigate
the consistency between theoretically predicted power-laws and those
seen experimentally.

{\it Acknowledgements.---}This work was supported by the Royal Society
and the EPSRC under grant number EP/D036194/1. This work was
motivated by early conversations with Prof. C. Pepin during a visit
to Saclay. We would also like to thank Dr. S. A. Grigera,
Dr. C. A. Hooley, and Prof. A. P. Mackenzie for comments on the
manuscript.

\section{Appendix A}
In this appendix, we outline the evaluation of the scattering rates
given in Eq.(\ref{timescales}). The details of the calculation is somewhat similar to that of
the same relaxation rate due to phonon scattering. Because of this
similarity, our analysis follows quite closely --- and makes explicit
reference to --- the calculation of electron-phonon relaxation rates
presented in Chapter 8 of the textbook of Mahan\cite{Mahan}. We shall carry out the calculations in $d$-dimensions, where $d = 2$ or $3$ in the physical system.

Our first task is to make a few manipulations of the energy relaxation
rate to put it in a simplified form. The first step involves using the detailed
balance relation expressed in the form
$$
\gamma_{{\bm{p}}{\bm{q}}}
=
\gamma_{{\bm{q}}{\bm{p}}}
\frac{f^T_{\bm{q}}(1-f^T_{\bm{p}})}{f^T_{\bm{p}}(1-f^T_{\bm{q}})}
=
\gamma_{{\bm{q}}{\bm{p}}}
\frac{n^T(\Delta \epsilon_{\bm{q}})}{n^T(\Delta \epsilon_{\bm{q}})+1},
$$
where
$n^T(\Delta \epsilon_{\bm{q}})$
is the Bose distribution at temperature $T$ and
$\Delta \epsilon_{\bm{q}}=\epsilon_{\bm{q}}-\epsilon_{\bm{p}}$.
The next step is to
integrate over $|{\bm{q}}|$ assuming that the $|{\bm{q}}|$-dependence of
terms other than the Fermi-distribution functions is small and also
that the density of electronic states is constant at the Fermi
surface. In so doing, we encounter two integrals
\begin{eqnarray*}
I_1(\omega)
&=&
\int d \epsilon f(\epsilon) (1-f(\epsilon -\omega))
=
\omega n(\omega),
\\
I_2(\omega)
&=&
\int d \epsilon f(\epsilon-\omega) (1-f(\epsilon))
=
\omega (n^0(\omega)+1).
\end{eqnarray*}
After these manipulations, the energy relaxation may be expressed as
$$
\frac{d {\cal E}}{dt}
=
\frac{1}{2}
\int \frac{d^d {\bm{p}} d^d \hat {\bm{q}}}{(2 \pi)^{2d}}
 \rho_{\textrm{F}}
\frac{
\gamma_{{\bm{q}}{\bm{p}}}\Delta \epsilon_{\bm{q}}^2
}{
n^T(\Delta \epsilon_{\bm{q}})+1}
\left[
n^0(\Delta \epsilon_{\bm{q}}) - n^T(\Delta \epsilon_{\bm{q}})
\right],
$$
where $\rho_{\textrm{F}}$ is the electronic density of states at the Fermi
surface and the remaining integral over ${\bm{q}}$ is just angular; functions of ${\bm{q}}$ are to be interpreted as having $|{\bm{q}}|$ equal to the Fermi wavevector.
The superscripts $0$ and $T$ on the Bose-distribution function indicate that they are at the base
temperature or the elevated temperature, $T$, respectively.
Notice that this is automatically zero when $T^0=T$.

In the limit $T^0 \rightarrow 0$, we may neglect $n^0(\Delta \epsilon_{\bm{q}})$. Also, in the limit where $\Delta \epsilon_{\bm{q}} \ll T$, we may write the energy relaxation in the same form as the other relaxation rates using $n^T(\Delta \epsilon_{\bm{q}})/(n^T(\Delta \epsilon_{\bm{q}})+1) \simeq (1-f^T_{\bm{p}})/(1-f^T_{\bm{q}})$. As all the terms in our scattering rates are now at the elevated temperature $T$, we will from now on omit this superscript for clarity.

So far, we have reduced our scattering rates to the forms
\begin{eqnarray}
\frac{1}{\tau_{\bm{q}}}
&=&
\gamma_{\bm{q}}
=
\int \frac{d^d {\bm{p}}}{(2 \pi)^d}
\gamma_{{\bm{q}}{\bm{p}}}
\frac{1-f_{\bm{p}}}{1-f_{\bm{q}}},
\nonumber\\
\frac{1}{\tau^{\textrm{tr}}_{\bm{q}}}
&=&
\gamma^{\textrm{tr}}_{\bm{q}}
=
\int \frac{d^d {\bm{p}}}{(2 \pi)^d}
\gamma_{{\bm{q}}{\bm{p}}}
\frac{1-f_{\bm{p}}}{1-f_{\bm{q}}}
\left[
1
-
\frac{{\bm{q}}.{\bm{p}}}{{\bm{q}}^2}
\frac{\gamma_{\bm{q}}^{\textrm{tr}}}{\gamma_{\bm{p}}^{\textrm{tr}}}
\right],
\nonumber\\
\frac{d {\cal E}}{dt}
&=&
-
\frac{1}{2}
\int \frac{d^d {\bm{p}}}{(2 \pi)^d}\frac{d^d \hat {\bm{q}}}{(2 \pi)^d}
 \rho_{\textrm{F}}
\Delta \epsilon_{\bm{q}}^2
\gamma_{{\bm{q}}{\bm{p}}}
\frac{
1-f_{\bm{p}}
}{
1-f_{\bm{q}}
}.
\label{full_relaxations}
\end{eqnarray}
In the present case, we are
interested in scattering from critical magnons.
$\gamma_{{\bm{p}}{\bm{q}}}$ then takes the form given by
Eq.~(\ref{MagnonGamma}).
In order to calculate the various relaxation rates explicitly, it is
useful to introduce a generalization of the McMillan function. In the discussion of electron-phonon
scattering, this takes the form\cite{Mahan}
\begin{equation}
\alpha^2 F(E,\omega)
=
\frac{\hbar}{2 \pi}
\int
\frac{d^d {\bm{q}}}{(2 \pi)^d}
|g_{\bm{q}}|^2
\delta(\omega-\omega_{\bm{q}})
\delta (E- \epsilon_{{\bm{k}}+{\bm{q}}}),
\label{McMillan_Mahan}
\end{equation}
where $\omega_{\bm{q}}$ is the frequency of a phonon with momentum $\bm{q}$ and $\epsilon_{\bm{p}}$ is the energy of an electron with momentum
${\bm{p}}$. The generalisation of this to scattering from overdamped
modes is given by
\begin{equation}
\alpha^2 F(E,\omega)
=
\frac{\hbar}{2 \pi}
\int
\frac{d^d {\bm{q}}}{(2 \pi)^d}
|g_{\bm{q}}|^2
\rho({\bm{q}},\omega)
\delta (E- \epsilon_{{\bm{k}}+{\bm{q}}}),
\label{McMillan_Mahan1}
\end{equation}
where $\rho({\bm{q}}, \omega) = {\cal I}m D^{\textrm{R}}({\bm{q}},\omega)$ is the
magnon spectral function. The analogous McMillan function for
transport is given by
\begin{equation}
\alpha_t^2 F(E,\omega)
=
-
\frac{\hbar}{2 \pi}
\int
\frac{d^d {\bm{q}}}{(2 \pi)^d}
|g_{\bm{q}}|^2
\frac{{\bm{q}}.{\bm{k}}}{{\bm{k}}^2}
\rho({\bm{q}},\omega)
\delta (E- \epsilon_{{\bm{k}}+{\bm{q}}}).
\label{McMillan_Mahan2}
\end{equation}
The additional angular factors weight scattering at different
angles in the usual
way. Eqs.~(\ref{McMillan_Mahan1},\ref{McMillan_Mahan2}) are the
generalisations of Eqs.(8.145,8.146) in Ref.\cite{Mahan}.

With this identification, expressions for the various scattering integrals may be obtained in the limit $\Delta \epsilon_{\bm{q}} \ll T$ as follows (see for example Section 8.3.1 of Ref.\cite{Mahan} --- we use $(1-f_{\bm{p}})/(1-f_{\bm{q}}) \approx n(\Delta \epsilon_{\bm{q}})$):
\begin{widetext}
\begin{eqnarray}
\frac{1}{\tau_{\bm{k}}}
&=&
2\frac{2 \pi}{\hbar}
\int^{\infty}_0  \! d \omega n(\omega)
\alpha^2 F(\epsilon_{\bm{k}},\omega)
=
2
\int^{\infty}_0  \! d \omega
\frac{d^d {\bm{q}}}{(2 \pi)^d}
|g_{\bm{q}}|^2
n(\omega)
\rho({\bm{q}},\omega)
\delta (\epsilon_{{\bm{k}}+{\bm{q}}}-\epsilon_{{\bm{k}}}),
\label{McMillan1}
\\
\frac{1}{\tau^{\textrm{tr}}_{\bm{k}}}
&=&
2\frac{2 \pi}{\hbar}
\int^{\infty}_0 \!\! d \omega n({\omega})
\alpha_t^2 F(\epsilon_{\bm{k}},\omega)
=
-
2
\int^{\infty}_0 \!\! d \omega
\frac{d^d {\bm{q}}}{(2 \pi)^d}
|g_{\bm{q}}|^2
n(\omega)
\frac{{\bm{q}}.{\bm{k}}}{{\bm{k}}^2}
\rho({\bm{q}},\omega)
\delta (\epsilon_{{\bm{k}}+{\bm{q}}}-\epsilon_{{\bm{k}}}),
\label{McMillan2}
\\
\frac{d {\cal E}}{dt}
&=&
4 \pi \hbar^2 \rho_{\textrm{F}}
\int^{\infty}_0 d \omega n(\omega) \omega^2
\alpha^2 F(\epsilon_{\bm{k}},\omega)
=
2 \hbar^3\rho_{\textrm{F}}
\int^{\infty}_0 d \omega
\frac{d^d {\bm{q}}}{(2 \pi)^d}
|g_{\bm{q}}|^2
n(\omega) \omega^2
\rho({\bm{q}},\omega)
\delta (\epsilon_{{\bm{k}}+{\bm{q}}}-\epsilon_{{\bm{k}}}).
\label{McMillan3}
\end{eqnarray}
\end{widetext}
These integrals may be simplified by first linearising the electron
energy near to the Fermi surface;
$\epsilon_{{\bm{k}}+{\bm{q}}}-\epsilon_{\bm{k}}
\approx
{\bm{v}}_{\bm{k}}.{\bm{q}}
=
v_{\textrm{F}} \cos \theta |{\bm{q}}|
$, where ${\bm{v}}_{\bm{k}}$ is the Fermi velocity and $\theta $ is the
angle between $\bm{k}$ and $\bm{q}$. Assuming that the matrix element
does not have a significant angular dependence---
an assumption
  that is only true for ferromagnets; antiferromagnets have hot lines
  of scattering on the Fermi surface where electron states are related
  by the ordering wave-vector that lead to complications in this latter case\cite{Rosch}---
 the angular integrals
over $\bm{q}$ may then be carried out. One then obtains
\begin{eqnarray}
\frac{1}{\tau_{\bm{k}}}
&=&
\frac{2  g^2}{(2 \pi)^2 v_{\textrm{F}}}
\int^{\infty}_0 d \omega n(\omega)
\nonumber\\
& &
\;\;\;\;\;\;\;\;\;\;\;\;\;\;\;\;\;\;\;\;\;\;
\times
\int^{T/v_{\textrm{F}}}
d| {\bm{q}}| |{\bm{q}}|^{d-2}
\rho({\bm{q}},\omega)
\nonumber\\
\frac{1}{\tau^{\textrm{tr}}_{\bm{k}}}
&=&
\frac{2  g^2}{(2 \pi)^2 v_{\textrm{F}}}
\frac{1}{2 k_{\textrm{F}}^2}
\int^{\infty}_0 d \omega n(\omega)
\nonumber\\
& &
\;\;\;\;\;\;\;\;\;\;\;\;\;\;\;\;\;\;\;\;\;\;
\times
\int^{T/v_{\textrm{F}}} d| {\bm{q}}| |{\bm{q}}|^d
\rho({\bm{q}},\omega)
\nonumber\\
\frac{d {\cal E}}{dt}
&=&
 \hbar^3 \rho_{\textrm{F}}
\frac{ 2  g^2}
{(2 \pi)^2 v_{\textrm{F}}}
\int^{\infty}_0 d \omega n(\omega) \omega^2
\nonumber\\
& &
\;\;\;\;\;\;\;\;\;\;\;\;\;\;\;\;\;\;\;\;\;\;
\times
\int^{T/v_{\textrm{F}}} d| {\bm{q}}| |{\bm{q}}|^{d-2}
\rho({\bm{q}},\omega)
\nonumber\\
\end{eqnarray}
The momentum integral has acquired an explicit cut-off at $T/v_{\textrm{F}}$ after
linearising the electron energy at the Fermi surface. The remaining
integrals may be calculated after
explicit substitution of the magnon spectral function.

We will
evaluate these expressions in both high- and low-temperature
limits. Which of these limits one is in is determined by a comparison of $r(T)$  with the typical value of $|
{\bm{q}}|^2$. The former varies with temperature according to $r(T) \sim T^{\frac{d+z-2}{z}}$ and the latter
as $T^2$. In the low-temperature limit, $r \gg {\bm{q}}^2$ and in the high-temperature limit $r \ll {\bm{q}}^2$.
In both cases, we consider temperatures much less than the Fermi energy, $T\ll \epsilon_{\textrm{F}}$.

Carrying out the integrals in the {\it low-temperature} limit, we find
\begin{eqnarray*}
\frac{1}{\tau_{\bm{k}}}
&=&
\frac{2  g^2}{(2 \pi)^2 v_{\textrm{F}}}
\int^{\infty}_0 d \omega \; n(\omega)
\int^{\frac{T}{v_{\textrm{F}}}}_0 d| {\bm{q}}| |{\bm{q}}|^{d-2}
\rho({\bm{q}},\omega)
\nonumber\\
&\sim&
\int^{\frac{T}{v_{\textrm{F}}}}_0 d| {\bm{q}}| |{\bm{q}}|^{d-2}
\int^{\infty}_0 d \omega  \; n(\omega)
\frac{\omega/ \Gamma_{\bm{q}}}
{r^2+(\omega/\Gamma_{\bm{q}})^2}
\nonumber\\
&\sim&
\int^{\frac{T}{v_{\textrm{F}}}}_0 d| {\bm{q}}| |{\bm{q}}|^{d-2}\Gamma_{\bm{q}}
\int^{\infty}_0 \frac{d \omega}{r\Gamma_{\bm{q}}}  \; n(\omega)
\frac{\omega/ \Gamma_{\bm{q}}}
{1+(\omega/r\Gamma_{\bm{q}})^2}
\nonumber\\
&\sim&
\int^{\frac{T}{v_{\textrm{F}}}}_0 d| {\bm{q}}| |{\bm{q}}|^{d-2}\Gamma_{\bm{q}}
\int^{\infty}_0 du  \; n(u r \Gamma_{\bm{q}})
\frac{u}
{1+u^2}
\nonumber\\
&\sim&
\frac{T}{r(T)}
\int^{\frac{T}{v_{\textrm{F}}}}_0 d| {\bm{q}}| |{\bm{q}}|^{d-2}
\int^{\infty}_0 du  \;
\frac{1}
{1+u^2}
\nonumber\\
&\sim&
\frac{T^d}{r(T)},
\end{eqnarray*}
where we have used the fact that $r \gg {\bm{q}}^2$ at low
temperatures. The frequency integral is dominated by the region where
$\omega$ is up to order $ r(T) \Gamma_{\bm{q}}$. For ${\bm{q}} \sim T$,
at these frequencies $\omega/T \sim \Gamma r(T)T^{z-3} \ll 1$ and the
Bose distribution can be approximated by its low-frequency limit $n(x)
\sim T/x$. As a final consistency check, we need to make sure that the
dominant momentum ${\bm{q}}$ is of the order of $T$ as indeed it is.

A similar evaluation of the transport scattering rate
yields the result
\begin{equation}
\frac{1}{\tau^{\textrm{tr}}_{\bm{k}}}
\sim
\frac{T^{d+2}}{r(T)}
.
\end{equation}
The energy relaxation is given by
\begin{eqnarray}
\frac{d {\cal E}}{dt}
&=&
 \hbar^3 \rho_{\textrm{F}}
\frac{ 2  g^2}
{(2 \pi)^2 v_{\textrm{F}}}
\int^{\infty}_0
\!\!\!
d \omega \; \omega^2 n(\omega)
\int^{\frac{T}{v_{\textrm{F}}}}_0
\!\!\!
d| {\bm{q}}|  |{\bm{q}}|^{d-2}
\rho({\bm{q}},\omega)
\nonumber\\
&\sim&
\int^{\frac{T}{v_{\textrm{F}}}}_0
\!\!\!
d| {\bm{q}}|  |{\bm{q}}|^{d-2}
\int^{\infty}_0
\!\!\!
d \omega \; \omega^2 n(\omega)
\frac{\omega/ \Gamma_{\bm{q}}}
{r^2+(\omega/\Gamma_{\bm{q}})^2}
\nonumber\\
&\sim&
r T
\int^{\frac{T}{v_{\textrm{F}}}}_0
\!\!\!
d| {\bm{q}}|  |{\bm{q}}|^{d-2} \Gamma_{\bm{q}}^2
\int^{T/r \Gamma_{\bm{q}} }_0
\!\!\!
d u
\frac{u^2}{1+u^2}
\nonumber\\
&\sim&
r T
\int^{\frac{T}{v_{\textrm{F}}}}_0
\!\!\!
d| {\bm{q}}|  |{\bm{q}}|^{d-2} \Gamma_{\bm{q}}^2
\left[\frac{T}{r \Gamma_{\bm{q}}}-\frac{\pi}{2} \right]
\nonumber\\
&\sim&
 T^2
\int^{\frac{T}{v_{\textrm{F}}}}_0
\!\!\!
d| {\bm{q}}|  |{\bm{q}}|^{d-2} \Gamma_{\bm{q}}
\nonumber\\
&\sim&
 T^2
\int^{\frac{T}{v_{\textrm{F}}}}_0
\!\!\!
d| {\bm{q}}|  |{\bm{q}}|^{d+z-4}
\nonumber\\
&\sim&
 T^{d+z-1}
.
\end{eqnarray}

Carrying out the same integrations in the {\it high-temperature}
limit, we find
\begin{eqnarray}
\frac{1}{\tau_{\bm{k}}}
&=&
\frac{2  g^2}{(2 \pi)^2 v_{\textrm{F}}}
\int^{\infty}_0 d \omega n(\omega)
\int^{\frac{T}{v_{\textrm{F}}}}_0 d| {\bm{q}}| |{\bm{q}}|^{d-2}
\rho({\bm{q}}, \omega)
\nonumber\\
&\sim&
\int^{\infty}_0 d \omega n(\omega)
\int^{\frac{T}{v_{\textrm{F}}}}_0 d| {\bm{q}}| |{\bm{q}}|^{d-2}
\frac{\omega/\Gamma |{\bm{q}}|^{z-2} }{
{\bm{q}}^4+(\omega/\Gamma |{\bm{q}}|^{z-2})^2}
\nonumber\\
&\sim&
\int^{\infty}_0 d \omega n(\omega)
\int^{\frac{T}{v_{\textrm{F}}}}_0 d| {\bm{q}}| |{\bm{q}}|^{d+z-4}
\frac{\omega
}{
\Gamma^2 {\bm{q}}^{2z}+\omega^2
}
\nonumber\\
&\sim&
\int^{\infty}_0 d \omega \frac{n(\omega)}{\omega}
\int^{\frac{T}{v_{\textrm{F}}}}_0 d| {\bm{q}}|
\frac{|{\bm{q}}|^{d+z-4}
}{
\Gamma^2 {\bm{q}}^{2z}/\omega^2+1
}
\nonumber\\
&\sim&
\int^{\infty}_0 d \omega n(\omega)
\omega^{\frac{d-3}{z}}
\int^{\frac{T}{v_{\textrm{F}}} \left( \frac{\Gamma}{\omega} \right)^{1/z}}_0 du
\frac{u^{d+z-4}
}{
u^{2z}+1
}
\nonumber\\
&\sim&
T^{\frac{d+z-3}{z}}
\int^{\infty}_0 d v \;n(vT)
v^{\frac{d-3}{z}}
\int^{\infty}_0 du
\frac{u^{d+z-4}
}{
u^{2z}+1
}
\nonumber\\
& \sim&
T^{\frac{d+z-3}{z}}
\end{eqnarray}
In carrying out these manipulations we have used the fact that
$\frac{T}{v_{\textrm{F}}} \left( \frac{\Gamma}{\omega} \right)^{1/z}
\rightarrow \infty$
at high temperatures, which is consistent since the dominant
contribution to the frequency integral comes from $\omega \sim T$.
We have rescaled the momentum and frequency integrals and then  and
used the explicit substitution
$\Gamma_{\bm{q}}=\Gamma |{\bm{q}}|^{z-2}$. A similar evaluation
of the transport scattering rate yields
\begin{equation}
\frac{1}{\tau_{\bm{k}}^{\textrm{tr}}}
\sim
T^{\frac{d+z-1}{z}},
\end{equation}
{\it i.e.} it carries an extra factor of ${\bm{q}}^2 \sim T^{2/z}$
compared with the scattering rate. Finally, the energy relaxation rate
is given by
\begin{eqnarray}
\frac{d {\cal E}}{dt}
&=&
 \hbar^3 \rho_{\textrm{F}}
\frac{ 2 g^2}
{(2 \pi)^2 v_{\textrm{F}}}
\int^{\infty}_0 d \omega \; n(\omega) \omega^2
\nonumber\\
& &
\;\;\;\;\;\;\;\;\;\;\;\;\;\;\;\;\;\;\;\;\;\;\;
\times
\int^{\frac{T}{v_{\textrm{F}}}}_0 d| {\bm{q}}|  |{\bm{q}}|^{d-2}
\rho({\bm{q}},\omega)
\nonumber\\
&\sim&
\int^{\infty}_0 d \omega n(\omega) \omega^2
\int^{\frac{T}{v_{\textrm{F}}}}_0 d| {\bm{q}}| |{\bm{q}}|^{d+z-4}
\frac{\omega
}{
\Gamma^2 {\bm{q}}^{2z}+\omega^2
}
\nonumber\\
&\sim&
\int^{\infty}_0 d \omega n(\omega) \omega
\int^{\frac{T}{v_{\textrm{F}}}}_0 d| {\bm{q}}|
\frac{|{\bm{q}}|^{d+z-4}
}{
\Gamma^2 {\bm{q}}^{2z}/\omega^2+1
}
\nonumber\\
&\sim&
\int^{\infty}_0 d \omega n(\omega)
\omega^{\frac{d+2z-3}{z}}
\int^{\frac{T}{v_{\textrm{F}}} \left( \frac{\Gamma}{\omega} \right)^{1/z}}_0 du
\frac{u^{d+z-4}
}{
u^{2z}+1
}
\nonumber\\
&\sim&
T^{\frac{d+3z-3}{z}}
\int^{\infty}_0 d v \;n(vT)
v^{\frac{d+2z-3}{z}}
\int^{\infty}_0 du
\frac{u^{d+z-4}
}{
u^{2z}+1
}
\nonumber\\
& \sim&
T^{\frac{d+3z-3}{z}}
\end{eqnarray}



\begin{thebibliography}{99}
\bibitem{Sachdev} S. Sachdev, {\it Quantum Phase Transitions} (CUP,
  Cambridge, 1999).

\bibitem{Sondhi97} S. L. Sondhi, S. M. Girvin, J. P. Carini, and
  D. Shahar, Rev. Mod. Phys. {\bf 69}, 315 (1997).

\bibitem{Coleman05} P. Coleman and A. J. Schofield, Nature {\bf 433},
  226 (2005).

\bibitem{Moriya} T. Moriya, {\it Spin fluctuations in itinerant
    electron magnetism} (Springer-Verlag, Berlin, 1985).

\bibitem{Hertz76} J. A. Hertz, Phys. Rev. B {\bf 14}, 1165 (1976).

\bibitem{Millis} A. J. Millis, Phys. Rev. B {\bf 48}, 7183 (1993).


\bibitem{Dalidovich04} D. Dalidovich and P. Phillips, Phys. Rev. Lett. {\bf 93}, 027004 (2004).

\bibitem{Green05} A. G. Green and S. L. Sondhi, Phys. Rev. Lett. {\bf 95}, 267001 (2005).

\bibitem{Green06} A. G. Green, J. E. Moore, S. L. Sondhi, and A. Vishwanath, Phys. Rev. Lett. {\bf 97}, 227003 (2006).

\bibitem{Damle97} K. Damle and S. Sachdev, Phys. Rev. B {\bf 56}, 8714 (1997).

\bibitem{Cha91} M.-C. Cha, M. P. A. Fisher, S. M. Girvin, M. Wallin, and A. P. Young, Phys. Rev. B {\bf
44}, 6883 (1991).

\bibitem{Fenton05} Non-linear response near to quantum criticality has also been discussed in J. Fenton
and A. J. Schofield, Phys. Rev. Lett. {\bf 95}, 247201 (2005), although in their case, the system remains
in thermal equilibrium.

\bibitem{Aoki} T. Oka and H. Aoki, Phys. Rev. Lett. {\bf 95}, 137601 (2005).

\bibitem{footnote1} In a previous work considered by one of us\cite{Green05}, the
  situation was rather simpler since the thermal conductivity of the
  model system considered was formally infinite and therefore there
  was no limit to the rate at which Joule heat could be transported.

\bibitem{Mitra06} A. Mitra, S. Takei, Y. B. Kim, and A. J. Millis, Phys. Rev. Lett. {\bf 97},
236808 (2006).

\bibitem{footnote2} The relation that the present work bears to
  Ref.\cite{Mitra06}. is rather similar to that of
  Ref.\cite{Green05} to Ref.\cite{Dalidovich04}.

\bibitem{footnote3}
The expression for the transport relaxation rate is arrived at by
first substituting the form
$\delta f_{\bm{p}}=g(|{\bm{p}}|) {\bm{p}}.{\bm{E}}$
into the scattering integral.
\begin{eqnarray*}
& &
\gamma_{\bm{q}}
[1-M]_{{\bm{q}}{\bm{p}}} \delta f_{\bm{p}}
\nonumber\\
&=&
\int \frac{d {\bm{p}}}{(2 \pi)^3}
\left[
\gamma_{{\bm{q}}{\bm{p}}}
\frac{1-f_{\bm{p}}}{1-f_{\bm{q}}}
g(|{\bm{q}}|) {\bm{q}}.{\bm{E}}
-
\gamma_{{\bm{p}}{\bm{q}}}
\frac{1-f_{\bm{q}}}{1-f_{\bm{p}}}
g(|{\bm{p}}|) {\bm{p}}.{\bm{E}}
\right]
\nonumber\\
&=&
\int \frac{d {\bm{p}}}{(2 \pi)^3}
\gamma_{{\bm{q}}{\bm{p}}}
\frac{1-f_{\bm{p}}}{1-f_{\bm{q}}}
\left[
g(|{\bm{q}}|) {\bm{q}}.{\bm{E}}
-
\frac{f_{\bm{q}}(1-f_{\bm{q}})}{f_{\bm{p}}(1-f_{\bm{p}})}
g(|{\bm{p}}|) {\bm{p}}.{\bm{E}}
\right]
\nonumber\\
&=&
\int \frac{d {\bm{p}}}{(2 \pi)^3}
\gamma_{{\bm{q}}{\bm{p}}}
\frac{1-f_{\bm{p}}}{1-f_{\bm{q}}}
\left[
g(|{\bm{q}}|) {\bm{q}}.{\bm{E}}
-
\frac{\partial_{\epsilon}f_{\bm{q}}}{\partial_{\epsilon} f_{\bm{p}}}
g(|{\bm{p}}|) {\bm{p}}.{\bm{E}}
\right]
\nonumber\\
&=&
\delta f_{\bm{q}}
\int \frac{d {\bm{p}}}{(2 \pi)^3}
\gamma_{{\bm{q}}{\bm{p}}}
\frac{1-f_{\bm{p}}}{1-f_{\bm{q}}}
\left[
1
-
\frac{\partial_{\epsilon}f_{\bm{q}}}{\partial_{\epsilon} f_{\bm{p}}}
\frac{g(|{\bm{q}}|){\bm{q}}.{\bm{E}}}{g(|{\bm{p}}|){\bm{p}}.{\bm{E}}}
\right]
\end{eqnarray*}
The final manipulation to get this to the same form as above to us the
fact that within the relaxation-time approximation
$g({\bm{p}})=\partial_{\epsilon} f_{\bm{p}}/ m \gamma^{\textrm{tr}}_{\bm{p}}$.

\bibitem{Tremblay79}
The heat sink is strictly necessary to
permit the formation of a steady state. It will not appear explicitly
in our analysis, which will be essentially linear response for the
electrons. The actual answer will turn out to be non-linear
in the electric field because of the field dependence of the
magnon-electron scattering rate. For an interesting discussion of the
role of the heat-sink in conductivity measurements see
A.-M. Tremblay, B. Patton, P. C. Martin, and P. F. Maldague, Phys. Rev. A {\bf 19}, 1721 (1979).

\bibitem{Lifshitz} E. M. Lifshitz and L. P. Pitaevskii, {\it Physical kinetics}, Section 82 (Course of Theoretical
Physics; Vol. 10, Butterworth-Heinemann, Oxford, 1981).

\bibitem{Kiess86} E. Kiess, Am. J. Phys. {\bf 55}, 1006 (1987).


\bibitem{Nagaosa} N. Nagaosa, {\it Quantum field theory in strongly correlated electronic systems}
(Springer-Verlag, Berlin, 1999).

\bibitem{Santiago} S.~A.~Grigera, private communication.

\bibitem{Mahan} G. D. Mahan, {\it Many-particle physics} (Kluwer Academic, New York, 2000).

\bibitem{Rosch} A. Rosch, Phys. Rev. Lett. {\bf 82}, 4280 (1999).

\end{thebibliography}
\end{document}